\documentclass[a4paper,aps,pra,superscriptaddress, reprint]{revtex4-2}
\usepackage[utf8]{inputenc}
\usepackage[T1]{fontenc}
\usepackage{amsmath, amsthm, amssymb,amsfonts,mathbbol,amstext}
\usepackage{graphicx}
\usepackage{dcolumn}
\usepackage{bm}
\usepackage{bbm}
\usepackage[usenames,dvipsnames]{xcolor}
\usepackage[]{hyperref}
\definecolor{C2}{RGB}{251, 77, 61}
\hypersetup{colorlinks=true, linkcolor=C2, citecolor=C2, urlcolor=C2}
\usepackage{mathtools}
\usepackage{comment}
\usepackage{color}
\usepackage{multirow}
\usepackage{pgf,tikz}
\usepackage{mathrsfs}
\usepackage{physics}
\usetikzlibrary{arrows}
\usepackage{xcolor}
\usepackage{soul}
\usepackage{adjustbox}
\usepackage{enumerate}
\usepackage[normalem]{ulem}
\usepackage{xcolor,cancel}
\usepackage{physics}

\newcommand{\ii}{\mathrm{i}}             

\begin{document}

\title{Reassessing thermodynamic advantage from indefinite causal order}

\author{Matheus Capela}
\affiliation{Department of Physical Chemistry, University of the Basque Country UPV/EHU, Apartado 644, 48080 Bilbao, Spain}
\affiliation{EHU Quantum Center, University of the Basque Country UPV/EHU}

\author{Harshit Verma}
\affiliation{Centre for Engineered Quantum Systems, School of Mathematics and Physics, The University of Queensland, St Lucia, QLD 4072, Australia}

\author{Fabio Costa}

\affiliation{Centre for Engineered Quantum Systems, School of Mathematics and Physics, The University of Queensland, St Lucia, QLD 4072, Australia}

\author{Lucas C. C\'{e}leri}
\email{lucas@qpequi.com}
\affiliation{QPequi Group, Institute of Physics, Federal University of Goi\'{a}s, 74.690-900, Goi\^{a}nia, Brazil}

\begin{abstract}
Indefinite causal order is a key feature involved in the study of quantum higher order transformations. Recently, intense research has been focused on possible advantages related to the lack of definite causal order of quantum processes. Quite often the quantum switch is claimed to provide advantages in information-theoretic and thermodynamic tasks. We address here the question whether indefinite causal order is a resource for quantum thermodynamics. Inspired by previous results in the literature, we show that indefinite causal order is not necessary for the reported increase in free energy and ergotropy. More specifically, we show that a simple causally ordered process, which replaces the system's state with a new one before the final measurement, outperforms the quantum switch in all thermodynamic tasks considered so far. We further show that a similar advantage can be also achieved without completely discarding system, if we allow for non-Markovian interactions between the system and an environment. We extend the analysis to more extreme examples of indefinite causal order, showing that they do not provide an advantage either. Finally, we discuss a possible way to study the advantages that may arise from indefinite causal order in a general scenario.
\end{abstract}
\maketitle

\section{Introduction}

Classical and quantum physics are causal in the sense that the relative temporal order between two given (causally connected) events are always defined~\cite{haag2012local}. However, there exist scenarios in which we can locally ascribe an indefinite causal order to quantum events. References~\cite{Hardy2005,Hardy2007} introduce indefinite causal order from the perspective of operational probability theories, while a model based on the structure of the Hilbert space is considered in Refs.~\cite{Chiribella2013,Oreshkov2012,Oreshkov2016}. Many theoretical and experimental developments followed these pioneering studies. Among these achievements, we can cite applications to thermodynamics~\cite{Dieguez2022,Felce2020,Guha2020,simonov2020ergotropy,Goldberg2021}, the quantum nature of gravity~\cite{Zych2019,parker2021background,baumann2022}, relativistic quantum information~\cite{foo2021,foo2020}, foundations of quantum mechanics~\cite{Shrapnel2017,Shrapnel2018causation,purves2021,milz2021,jia2018}, communication theory~\cite{Jia2019, Goswami2021,Loizeau2020}, quantum computation~\cite{araujo14, araujo2019}, quantum metrology~\cite{Blondeau2021,zhao2020}, and other information-theoretic tasks~\cite{Chiribella2012,feixquantum2015,Guerin2016}, just to mention a few recent ones. Recent experiments on these lines were also reported~\cite{cao2022,rubino2021,guo2020,procopio2019,wei2018} (see also the review~\cite{Goswami2020}).

A quantum process is said to have indefinite causal order if it cannot be written as a probabilistic mixture of processes with a fixed causal order, the so-called causally separable processes~\cite{Oreshkov2012,Araujo2015,Oreshkov2016b}. An interesting class of higher-order quantum operations violating this condition is the set of processes with quantum control of causal order~\cite{wechs2021quantum}, which includes the remarkable example of the quantum switch~\cite{Chiribella2013}. This class consists of processes where the order in which events occur is controlled by a quantum system, and it has been shown to be a valuable resource for various information-processing tasks~\cite{Chiribella2013,Chiribella2012,Guerin2016,Ebler2018,procopio2015,rubino2017,goswami2018,wei2019}. A resource theory approach for quantum control of causal orders was developed in Refs.~\cite{Taddei2019,Kristjansson2020}. 

The quantum switch was also employed in the context of quantum thermodynamics~\cite{Dieguez2022,Felce2020,Guha2020,simonov2020ergotropy}. In Ref.~\cite{Dieguez2022}, the quantum switch applied to two measurement channels was considered in the study of a thermal device (such as a heat engine, for instance). The authors of Ref.~\cite{Felce2020} claimed that the lack of causal order is responsible for an advantage in a refrigeration cycle over the ordered sequential use of quantum channels. In Ref.~\cite{Guha2020}, it has been considered that the thermalization process given by two distinct channels taking place in indefinite causal order provided by a quantum switch can enhance work extraction, when compared with the sequential version. In Ref.~\cite{simonov2020ergotropy}, the ergotropy was employed as a figure of merit in order to provide similar results as~\cite{Guha2020}. It is interesting to note that these results are based on the same fundamental task, the implementation of quantum control of causal order. It is important to mention that although the authors in Ref.~\cite{Guha2020} explicitly claim that the causal nonseparability of the process is responsible for the thermodynamic advantages found in their study, the same is not the case in Ref.~\cite{simonov2020ergotropy}.

The results in Refs.~\cite{Guha2020,simonov2020ergotropy} follow a similar methodology as Refs.~\cite{Ebler2018,Salek2018,Chiribella2021}, which show that the quantum switch can increase the communication capacity of noisy quantum channels. However, it has been demonstrated that simpler circuit models with \textit{definite} causal order can provide an even larger communication enhancement than the quantum switch, casting doubt on whether indefinite causal order offers any advantage for these particular tasks~\cite{guerin2019communication,abbott2020communication}. It is therefore reasonable to also ask to what extent indefinite causal order is responsible for the reported thermodynamic advantages.

In this work, we consider the role of processes with indefinite causal order in quantum thermodynamics. We address the question whether indefinite causal order implies any advantage for quantum thermodynamic tasks. Taking two figures of merit, in the same settings as in Refs.~\cite{Guha2020,simonov2020ergotropy} ---namely, free energy and ergotropy--- we show that indefinite causal order is not a fundamental resource for the tasks considered here, in a similar vein as Refs.~\cite{guerin2019communication,abbott2020communication}. Moreover, we show that non-Markovian quantum processes ---that is, causally ordered processes with environment-mediated temporal correlations--- imply thermodynamic advantages similar or beyond the quantum switch devices.

It is important to observe here that, instead of ruling out the possibility that the lack of causal order can be identified as a thermodynamic resource, our results indicate that a novel approach must be considered in order to identify such a contribution in a general scenario. We briefly discuss this issue at the end of the paper. 

The paper is organized as follows. In Sec.~\ref{processes} we review the process matrix formalism used throughout the text in order to represent the higher order quantum operations. The main results of the paper are presented in Section~\ref{results_main}, while a general discussion along with our final comments are left to Sec.~\ref{conclusions}. 

\section{Quantum processes} \label{processes}

To each quantum system $A$ it is associated a Hilbert space. We denote a quantum system and its Hilbert space with the same symbol. A composite quantum system is denoted as $AB$, for instance. The space of linear transformations from an input Hilbert space $A_I$ to an output Hilbert space $A_O$ is denoted as $L(A_I,A_O)$. In the case that the input and the output are equal, we make use of the notation $L(A) \coloneqq L(A,A)$. The state of a quantum system $A$ is a positive semi-definite operator $\rho \in L(A)$ with unit trace~\cite{watrous2018theory}.

We denote the collection of linear maps from $L(A_I)$ to $L(A_O)$ with $T(A_I,A_O)$. A quantum channel is a completely positive and trace-preserving linear map $\mathcal{A}\in~T(A_I,A_O)$~\cite{watrous2018theory}. 

There are several equivalent formulations of the process formalism. For our purposes, it is convenient to focus on the \emph{supermap} formulation, first introduced in Ref.~\cite{chiribella2008transforming}. The word supermap refers to the fact that it is a transformation of maps into maps. A supermap denoted with $\mathbf{W}$ is n-partite if it is a linear transformation of a collection of linear maps $(\mathcal{A}_{i})_{i=1}^{n}$ to a linear map $\mathcal{B}$. The input and output spaces of $\mathcal{B}$ are called the global past and future of the supermap, respectively. 

For instance, consider the case of bipartite supermaps, which take as input a pair of channels $\mathcal{A}_1 \in T(A_I^{(1)},A_O^{(1)})$ and $\mathcal{A}_2 \in T(A_I^{(2)},A_O^{(2)})$, and transform them into an output channel $\mathcal{B} \in T(P,F)$ with global past and future systems are denoted with $P$ and $F$, respectively. Processes are said to have the same form whenever they belong to the same class 
\begin{equation*}
    L(T(A_I^{(1)},A_O^{(1)}) \times T(A_I^{(2)},A_O^{(2)}),T(P,F)),
\end{equation*} 
or any isomorphic space; the symbol "$\times$" stands for the Cartesian product of sets. This definition extends trivially to general multipartite processes.

Consider a collection of bipartite quantum channels 
\begin{equation} \label{eq:bipartite_channels}
    \mathcal{A}_j \colon {L(A_I^{(j)} \tilde{A}_I^{(j)}) \rightarrow L(A_O^{(j)} \tilde{A}_O^{(j)})}, 
\end{equation}
with $j \in \{1,\dots,n\}$, where $I$ and $O$ labels input and output systems, respectively. A supermap acts trivially on the ancillary systems $\tilde{A}_I^{(j)},\,\tilde{A}_O^{(j)}$ if it has the form $\mathbf{W} \otimes \mathbf{I}_1 \otimes \cdots \otimes \mathbf{I}_n$, with $\mathbf{I}_i$ being the identity supermap acting on the space of linear maps $ L(\tilde{A}_I^{(i)}) \rightarrow L(\tilde{A}_O^{(i)})$. Importantly, we say that $\mathbf{W}$ is a valid quantum process whenever $\mathbf{W} \otimes \mathbf{I}_1 \otimes \cdots \otimes \mathbf{I}_n$ maps any collection of quantum channels $\mathcal{A}_j$ as defined in Eq.~(\ref{eq:bipartite_channels}) ---with input and output systems of arbitrary dimensions--- to an output quantum channel. We refer to Refs.~\cite{gour2019comparison,chiribella2008transforming} for a discussion on monopartite processes and consider here only the cases of bipartite and tripartite ones. The extension of the results presented here to the multipartite case is straightforward. 

\begin{figure}   
\begin{center}
\includegraphics[width=0.45\textwidth]{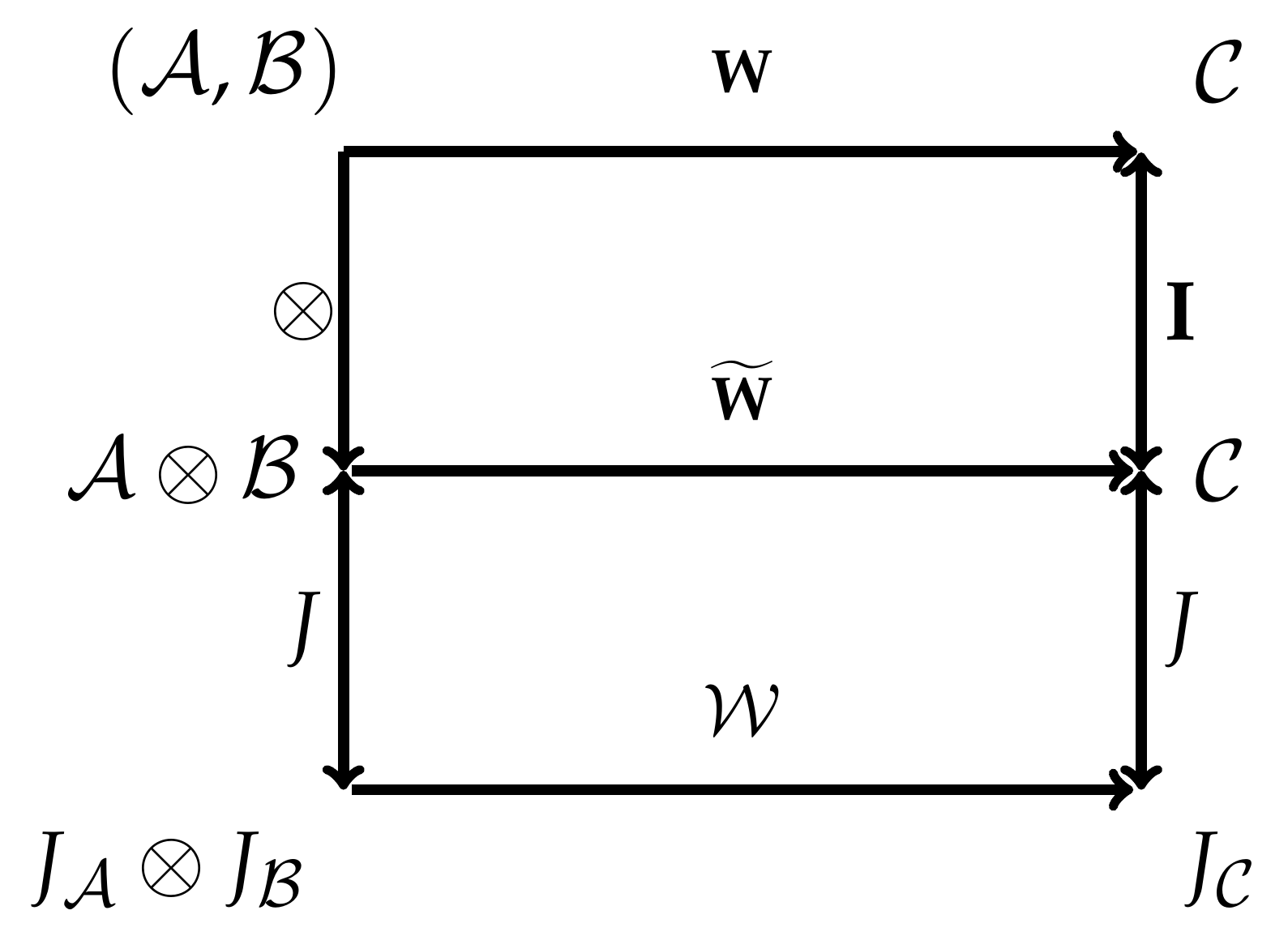}
\end{center}

    \caption{\textbf{Illustration of the Choi representation of a processes}. To each bilinear supermap $\mathbf{W}$, there is a unique linear supermap $\widetilde{\mathbf{W}}$ such that $\mathbf{W}(\mathcal{A},\mathcal{B}) = \widetilde{\mathbf{W}}(\mathcal{A} \otimes \mathcal{B})$. This property follows from the universal property of the tensor product~\cite{greub1978multilinear}.  Furthermore, those supermaps can be considered equivalent ones. The proof of existence of the linear map $\mathcal{W}$ mapping the tensor product of the Choi operators of the input channels $\mathcal{A}$ and $\mathcal{B}$ into the Choi operator of the output channel $\mathcal{C}$ can be found in Ref.~\cite{gour2019comparison}. The process matrix is defined as $\mathrm{W}=\mathrm{J}_{\mathcal{W}}$~\cite{Oreshkov2012,Araujo2015}.}
      \label{ChoiProcess}
\end{figure}

By means of the Choi representation of channels~\cite{choi1975completely,watrous2018theory}, it is possible to represent a process $\mathbf{W}$ in terms of an operator $\mathrm{W}$ called the process matrix of $\mathbf{W}$~\cite{chiribella2008transforming,Oreshkov2012}. Here, the Choi operator of a linear map $\mathcal{A}$ is denoted as $\mathrm{J}_{\mathcal{A}}$. Thus, if $\mathcal{C}=\mathbf{W}(\mathcal{A},\mathcal{B})$ (mapping $\mathcal{A}$ and $\mathcal{B}$ into $\mathcal{C}$), then it follows that
\begin{equation} \label{eq:process_matrix}
\mathrm{J}_{\mathcal{C}}=\Tr_{AB}[ \mathrm{W}^{T_{AB}} (\mathrm{J}_{\mathcal{A}} \otimes \mathrm{J}_{\mathcal{B}} \otimes \mathbb{1} )],
\end{equation}
where $A=A_IA_O$ (with equivalent definition for $B$), $T_{AB}$ is the partial transposition of subsystems $A$ and $B$, while $\mathbb{1} $ represents the identity operator acting on $C$. Figure~\ref{ChoiProcess} shows a diagram illustrating such a representation of quantum processes.
In order for a linear operator to represent the process matrix of a valid quantum process, it must satisfy specific conditions which are explicitly stated in Ref.~\cite{Araujo2015}.

\subsection{Relevant processes}

In this work, we will compare the performance of different processes for two thermodynamic tasks: free energy and ergotropy extraction. Our processes of interest are defined in what follows. The particular examples considered in this manuscript have been chosen in order to be compared with the previous results reported in Refs.~\cite{Guha2020,simonov2020ergotropy}.

\subsubsection{Channel composition and probabilistic mixtures}

\begin{figure}
\centering
\begin{minipage}[t]{0.5\textwidth}
\centering

\begin{center}
\includegraphics[width=0.9\textwidth]{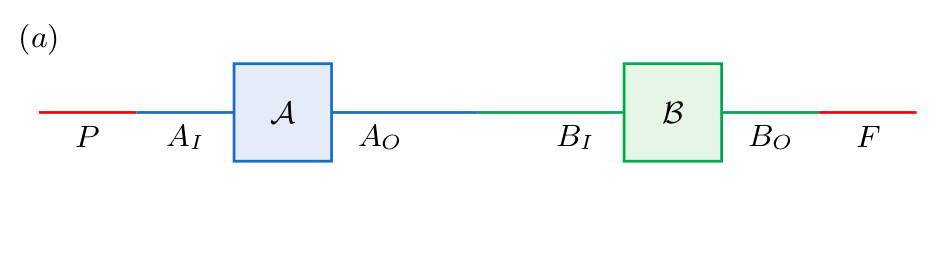}
\end{center}

\end{minipage}

\begin{minipage}{0.5\textwidth}
\centering

\begin{center}
\includegraphics[width=0.9\textwidth]{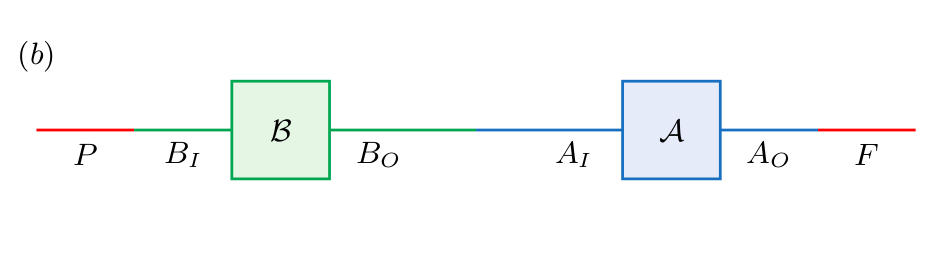}
\end{center}

\end{minipage}

\begin{minipage}{0.5\textwidth}
\centering

\begin{center}
\includegraphics[width=0.9\textwidth]{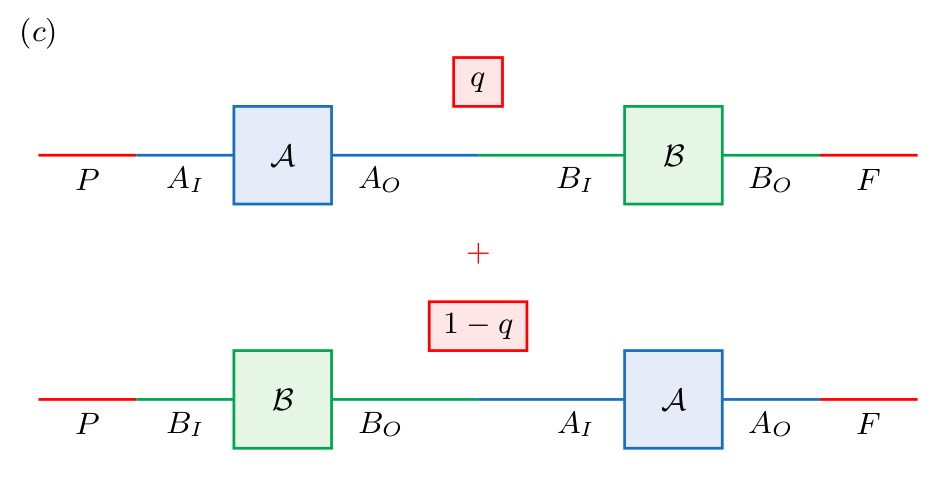}
\end{center}

\end{minipage}

\caption{\textbf{Probabilistic mixture of channel compositions.} The composition of channels $A \rightarrow B$ and $B \rightarrow A$ are diagrammatically represented in panels (a) and (b), respectively. Probabilistic mixtures of different orders of concatenation of channels are represented in panel (c).}
    \label{fig:CompositionProcess}
\end{figure}

Let us consider a simple bipartite process, which is given by a composition of two channels. There are two processes of this form, namely 
\begin{equation} \label{ABproc}
\mathbf{W}_{A \rightarrow B}(\mathcal{A},\mathcal{B}) = \mathcal{B}\circ\mathcal{A}
\end{equation}
and
\begin{equation} \label{BAproc}
\mathbf{W}_{B \rightarrow A} (\mathcal{A},\mathcal{B}) = \mathcal{A}\circ\mathcal{B}.
\end{equation}
In the first case, $A_O$ is a copy of $B_I$, while in the second one, $B_O$ is a copy of $A_I$. See Fig.~\ref{fig:CompositionProcess} for a pictorial representation.

Additionally, if all the involved spaces are copies of each other, we can define a probabilistic mixture of such a processes as
\begin{equation} \label{eq:composition_superchannel}
\mathbf{W}= q\mathbf{W}_{A \rightarrow B}+(1-q)\mathbf{W}_{B \rightarrow A},
\end{equation}
with $0 \leq q \leq 1$. 

Labelling the global input and output systems by $C_I$ and $C_O$, respectively, we can define the vector
\begin{equation} \label{ABvec}
\ket{A \rightarrow B}=| \mathbb{1} \rangle\rangle\otimes| \mathbb{1} \rangle\rangle\otimes| \mathbb{1} \rangle\rangle,
\end{equation}
where $| \mathbb{1} \rangle\rangle=\sum_{i}\ket{i}\otimes \ket{i}$ is the pure Choi representation of the identity operator for an orthonormal basis $\{\ket{i}\}$~\cite{Araujo2015}. The right-hand side of Eq.~\eqref{ABvec} is ordered as $PA_IA_OB_IB_OF$. 

Defining $\ket{B \rightarrow A}$ similar to Eq.~\eqref{ABvec}, but ordered as $PB_IB_OA_IA_OF$, we can construct the process matrices associated with $\mathbf{W}_{A \rightarrow B}$ and $\mathbf{W}_{B \rightarrow A}$ as
\begin{equation}
\mathrm{W}_{A\rightarrow B}=\ket{A \rightarrow B}\bra{A \rightarrow B} 
\end{equation}
and
\begin{equation}
\mathrm{W}_{B \rightarrow A}=\ket{B \rightarrow A}\bra{B \rightarrow A}, 
\end{equation}
respectively. A process is called pure if it is represented by a rank-1 process matrix $\mathrm{W}=\ket{w}\bra{w}$; in that case, $\ket{w}$ is called a process vector~\cite{Araujo2015}.

The process matrix associated with the process defined in Eq.~\eqref{eq:composition_superchannel} is given by
\begin{equation} \label{eq:composition_matrix}
\mathrm{W}=q\mathrm{W}_{A \rightarrow B}+(1-q)\mathrm{W}_{B \rightarrow A}. 
\end{equation}

Using Eq.~\eqref{eq:composition_matrix} along with Eq.~\eqref{eq:process_matrix}, we can construct a matrix representation of the process defined in Eq.~\eqref{eq:composition_superchannel}.

\subsubsection{Non-Markovian processes}

A Markov process is defined by the composition of quantum channels similarly to Eq.~(\ref{ABproc}), but allowing for arbitrary transformations on the input and output systems of the quantum channels $\mathcal{A}$ and $\mathcal{B}$.

\begin{figure}   
\begin{minipage}{0.5\textwidth}
\begin{center}
\includegraphics[width=0.9\textwidth]{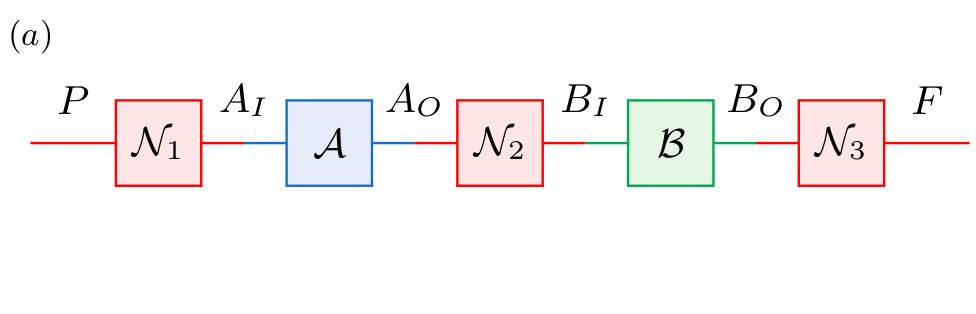}
\end{center}
\end{minipage}
\begin{minipage}{0.5\textwidth}
\begin{center}
\includegraphics[width=0.9\textwidth]{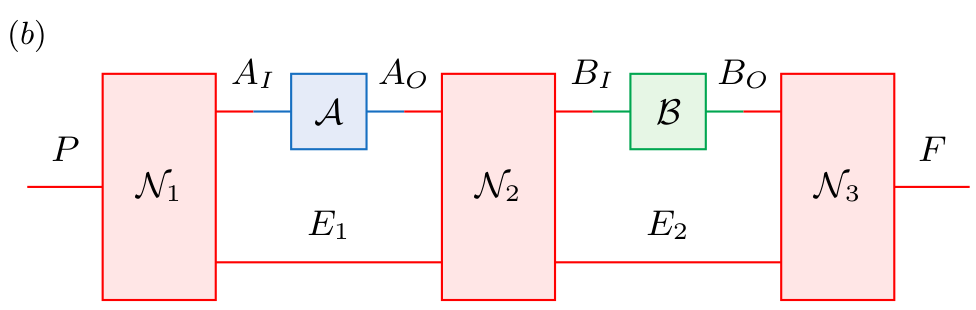}
\end{center}
\end{minipage}

\caption{\textbf{Causally ordered processes.} Markov processes are diagrammatically represented in top panel (a), while non-Markov processes -- allowing for quantum memory -- are represented in bottom panel (b).}

      \label{fig:Markov}
\end{figure}

Let $\mathcal{N}_{1}: L(P) \rightarrow L(A_I)$, $\mathcal{N}_{2}: L(A_O) \rightarrow L(B_I)$, and $\mathcal{N}_{3}: L(B_O) \rightarrow L(F)$ be quantum channels. A Markov process is a supermap of the form
\begin{equation} \label{markov}
    \mathbf{W}(\mathcal{A},\mathcal{B})=\mathcal{N}_{3} \circ \mathcal{B} \circ \mathcal{N}_{2} \circ \mathcal{A} \circ \mathcal{N}_{1}.
\end{equation}
We note that the composition process is recovered whenever $\mathcal{N}_{1}$, $\mathcal{N}_{2}$ and $\mathcal{N}_{3}$ are identity channels. See top panel of Fig.~\ref{fig:Markov} in comparison with the top panel of Fig.~\ref{fig:CompositionProcess}.

On the other hand, a process is non-Markovian if it cannot be written in the form~\eqref{markov}, which implies the presence of non-trivial system-environment correlations through a sequence of quantum operations~\cite{PhysRevLett.114.090402,Costa2016,pollock2018non,pollock2018operational,Shrapnel2018,Luchnikov2019, White2020,Milz2020kolmogorovextension,giarmatzi2021witnessing,White2021a}, see bottom panel of Fig.~\ref{fig:Markov}. We refer the reader to Refs.~\cite{milz2017introduction,PRXQuantum.2.030201} for review articles, and to Refs.~\cite{strasberg2019operational,strasberg2019repeated,strasberg2020thermodynamics} for applications in quantum stochastic thermodynamics. In general, a bipartite non-Markov quantum process is of the form
\begin{multline}\label{eq:non_Markov}
 \mathbf{W}_{A \prec B}(\mathcal{A},\mathcal{B}) = \\ \mathcal{N}_{3} \circ (\mathcal{B} \otimes \mathcal{I}^{E_2} ) \circ \mathcal{N}_{2} \circ (\mathcal{A} \otimes \mathcal{I}^{E_1}) \circ \mathcal{N}_{1}, 
\end{multline}
for appropriate fixed bipartite quantum channels $\mathcal{N}_{1}$, $\mathcal{N}_{2}$ and $\mathcal{N}_{3}$. The operation $\mathcal{I}^{E}$ is the identity channel acting on the environment $E$. Note that whenever $E_1$ and $E_2$ are trivial one-dimensional systems we recover Markov processes. 

In Eq.~\eqref{eq:non_Markov}, the quantum operation $\mathcal{A}$ precedes the operation $\mathcal{B}$. Nevertheless, we can define a non-Markovian process such that the opposite situation happens, denoted as $B \prec A$. A quantum process is called causally separable if it can be written as a probabilistic mixture of (possibly non-Markovian) processes with definite, albeit possibly different, causal orders~\cite{Oreshkov2012}. That is
\begin{equation}
    \mathbf{W}_{\text{sep}}= q  \mathbf{W}_{A \prec B} + (1-q)  \mathbf{W}_{B \prec A},
\end{equation}
with $0\leq q \leq 1$.

\subsubsection{Bipartite quantum switch}

A less trivial example of a process is the quantum control of causal order. The bipartite quantum switch was introduced in Ref.~\cite{Chiribella2012} in order to address the problem of quantum computation in processes with indefinite causal order.

The quantum switch has a bipartite global past and future denoted as $P=S_IQ_I$ and $F=S_OQ_O$, respectively. The ancilla $Q_{I(O)}$ is a qubit system responsible for controlling the two possible orderings, being therefore called the control system. On the other hand, the system $S_{I(O)}$ is called the target system. The input (output) target system $S_{I(O)}$ is a copy of the systems $A_{I(O)}$ and $B_{I(O)}$.

This is a pure process with associated process matrix denoted as $\mathrm{W}_{\textrm{2-switch}} = \ket{w_{\textrm{2-switch}}}\bra{w_{\textrm{2-switch}}}$, and a process vector given by two orthogonal terms
\begin{eqnarray}
 \ket{w_{\textrm{2-switch}}} &=& \ket{w_0} + \ket{w_1}. 
\end{eqnarray}

The vector defined as $\ket{w_0}=\ket{A \rightarrow B}\ket{0,0}$ is ordered as $S_IA_IA_OB_IB_OS_OQ_IQ_O$, while $\ket{w_1}=\ket{B \rightarrow A}\ket{1,1}$ is ordered according to $S_IB_IB_OA_IA_OS_OQ_IQ_O$. 

The quantum switch defines a quantum control of the orders $A \rightarrow B$ and $B \rightarrow A$. For instance, preparing the control system $Q_I$ in the state $\ket{0}$ results in the process
\begin{multline}
    \mathbf{W}_{\textrm{2-switch}}[\mathcal{A},\mathcal{B}](\rho \otimes \ket{0}\bra{0}) = \\ \mathbf{W}_{A \rightarrow B}[\mathcal{A},\mathcal{B}](\rho) \otimes \ket{0}\bra{0},
\end{multline}
while preparing it in the state $\ket{1}$ gives us 
\begin{multline}
    \mathbf{W}_{\textrm{2-switch}}[\mathcal{A},\mathcal{B}](\rho \otimes \ket{1}\bra{1}) = \\ \mathbf{W}_{B \rightarrow A}[\mathcal{A},\mathcal{B}](\rho) \otimes \ket{1}\bra{1}.
\end{multline}

Now, if the control system is started in the superposition $\ket{+} = (\ket{0}+\ket{1})/\sqrt{2}$ we have a non-trivial quantum control of the orders $A \rightarrow B$ and $B \rightarrow A$ We refer the reader to Ref.~\cite{Ebler2018,Araujo2015,branciard2016witnesses} for a detailed discussion of this phenomenon.

\subsubsection{Tripartite quantum switch}

When considering the sequential action of three channels $\mathcal{A}$, $\mathcal{B}$ and $\mathcal{C}$, there are 6 possible orderings, and moreover, we can also define a quantum control of the orders of three quantum operations. This quantum process is called tripartite quantum switch and the interested reader is referred to Refs.~\cite{procopio2020sending,wechs2021quantum} for more details.

Here, we consider a quantum control between the orders $A \rightarrow B \rightarrow C$ and $C \rightarrow B \rightarrow A$, which is given by the process vector
\begin{align} \label{eq:tripartite_switch}
\ket{w_{\textrm{3-switch}}} &= \frac{1}{\sqrt{2}}\left[\ket{A \rightarrow B \rightarrow C} \otimes \ket{0} \right. \nonumber \\ &\left. + \ket{C \rightarrow B \rightarrow A}  \otimes \ket{1}\right].
\end{align}
The vectors $\ket{A \rightarrow B \rightarrow C} \otimes \ket{0}$ and $\ket{C \rightarrow B \rightarrow A} \otimes  \ket{1}$ are defined as $| \mathbb{1} \rangle\rangle\otimes | \mathbb{1} \rangle\rangle \otimes | \mathbb{1} \rangle\rangle \otimes \ket{0}$ and $| \mathbb{1} \rangle\rangle\otimes | \mathbb{1} \rangle\rangle \otimes | \mathbb{1} \rangle\rangle \otimes \ket{1}$, with systems ordered as $S_IABCS_OQIQ_O$ and $S_I CBA S_O Q_IQ_O$ respectively. The systems $S_{I(O)}$ and $Q_{I(O)}$ are target and control systems, respectively. Thus, the process represented by $\mathrm{W}_{\textrm{3-switch}}=\ket{w_{\textrm{3-switch}}}\bra{w_{\textrm{3-switch}}}$ is tripartite, with global bipartite past and future $P=S_IQ_I$ and $F=S_OQ_O$, respectively.

\subsubsection{A tripartite process with indefinite causal order} \label{tripartite}

We consider the tripartite process $\mathbf{W}_{\textrm{Lugano}}$ defined in Ref.~\cite{araujo2017purification}. Each channel's input (output) system $A_{I(O)}$, $B_{I(O)}$, $C_{I(O)}$ is a qubit. The global past and future spaces, $P$ and $F$, consist of three qubits each, and therefore it can be written as a triple of qubits $P=P^{(1)} P^{(2)} P^{(3)}$ and $F=F^{(1)} F^{(2)} F^{(3)}$. 
This process has been shown to violate causal inequalities~\cite{baumeler2016space}, which implies indefinite causal order. It is interesting to observe here that the opposite is not true, since indefinite causal order does not necessarily implies a violation of a causal inequality. This is akin to a violation of a Bell inequality, which certainly implies entanglement, but the existence of entanglement is not enough for a state to violate a Bell inequality. It is defined by the process matrix
\begin{equation} \label{eq:det_process}
 W_{\textrm{Lugano}}=\ket{w_{\textrm{Lugano}}}\bra{w_{\textrm{Lugano}}}, 
\end{equation}
with the associated process vector
\begin{align} \label{eq:lugano}
 \ket{w_{\textrm{Lugano}}} & =  \sum_{i,j,k,r,s,t}\ket{r \oplus \lnot j \land k,s \oplus \lnot k \land i, t \oplus \lnot i \land j} \nonumber \\ & \otimes  \ket{r,s,t} \otimes \ket{i,j,k} \otimes  \ket{i,j,k} 
\end{align}
being ordered as $A_IB_IC_IPFA_OB_OC_O$. The summations run over $\{0,1\}$, while $\oplus$ represents addition modulo-$2$. The logical {\sc not} and {\sc and} operations are represented by $\lnot$ and $\land$, respectively. 

Some information related to previous results concerning the process defined in Eq.~\eqref{eq:lugano} are in order before we move to the next Section. The higher-order quantum operation $\mathbf{W}_{\textrm{Lugano}}$ was first defined in Ref.~\cite{baumeler2016space}, and shown there to be a valuable instance of a process violating causal inequalities. The superchannel version presented here appeared in Ref.~\cite{araujo2017purification}, where it has been shown to be an example of a process violating causal inequalities, and still, preserving unitary quantum operations. This process corresponds to a purely classical process with no causal order that can be interpreted as a closed time-like curve \cite{Baumeler2019, Tobar2020}. It has been called "Lugano process" in the literature; see Ref.~\cite{baumann2022noncausal}, for instance. 

\section{Quantum processes and thermodynamic tasks}\label{results_main}

We are now ready to present the main results of this study. This section is organized as follows. In Sec.~\ref{previous} we review the scenario in which indefinite causal order has been claimed to provide an advantage for quantum thermodynamics in Refs.~\cite{Guha2020,simonov2020ergotropy}. In Sec.~\ref{comparison} we present a criticism on the comparison of processes previously presented in literature. Section~\ref{results} contains our main results. In particular, we show that there is a causally ordered non-Markovian quantum process with similar performance as the quantum switch with respect to the thermodynamic tasks addressed in Refs.~\cite{Guha2020,simonov2020ergotropy}. Moreover, we show the existence of a quantum process with indefinite causal order for which no thermodynamic advantage can be extracted under the specified range of parameters considered in our protocol.

\subsection{Previous work}\label{previous}

We consider quantum processes acting on quantum channels whose input and output are qubit systems. The operations examined here are the Generalized Amplitude Damping (GAD) channel $\mathcal{R}_{p,\lambda}$~\cite{khatri2020information,wilde2011classical}, depending upon $0 \leq p \leq 1$ and $0 \leq \lambda \leq 1$, and the Phase Flip (PF) channel $\mathcal{T}_{q}$, parametrized by $0 \leq q \leq 1$~\cite{wilde2011classical}. The GAD channel has Kraus decomposition~\cite{Guha2020}
\begin{eqnarray}
    \mathrm{R}_{1}&=&\sqrt{p} ( \ket{0}\bra{0} + \sqrt{1-\lambda} \ket{1}\bra{1}), \\
    \mathrm{R}_{2}&=&\sqrt{1-p} (\sqrt{1-\lambda} \ket{0}\bra{0} + \ket{1}\bra{1}), \\
    \mathrm{R}_{3}&=&\sqrt{p \lambda} \ket{0}\bra{1}, \\
    \mathrm{R}_{4}&=&\sqrt{(1-p) \lambda } \ket{1}\bra{0}. \\
\end{eqnarray}
The GAD for $\lambda=1$ is denoted here as $\mathcal{R}_{p} \coloneqq \mathcal{R}_{p,1}$. The PF channel is represented by Kraus operators~\cite{Guha2020}
\begin{eqnarray}
    \mathrm{T}_{1} &=& \sqrt{q} \, \mathbb{1}, \\
    \mathrm{T}_{2} &=& \sqrt{1-q} \sigma_{z},
\end{eqnarray}
where $\sigma_{z}=\ket{0}\bra{0}-\ket{1}\bra{1}$ is the Pauli operator in $z$-direction.

The quantum operations $\mathcal{R}_{p,\lambda}(\cdot)=\sum_{i=1}^4 \textrm{R}_i(\cdot)\textrm{R}_i^{\dagger} $ and $\mathcal{T}_{q}(\cdot)=\sum_{i=1}^2 \textrm{T}_i(\cdot)\textrm{T}_i^{\dagger}$ are among the most studied noise models in the theory of open quantum systems~\cite{nielsen2000book}. Importantly, the composition of $\mathcal{R}_{p}$ and $\mathcal{T}_{q}$ in any order results in a thermalization process for input systems, resulting in a diagonal quantum state in the energy eigenbasis as the output. Let the initial system $S_I$ be in a diagonal state with respect to the energy eigenbasis, that is, $\rho=r\ket{0}\bra{0}+(1-r)\ket{1}\bra{1}$ with $0\leq r \leq 1$. Thus, any of the channel compositions $\mathcal{R}_{p} \circ \mathcal{T}_{q}$ and $\mathcal{T}_{q} \circ \mathcal{R}_{p}$ results in an output system $S_O$ in the thermal state~\cite{Guha2020}
\begin{equation}
    \tau=\frac{\exp(- \beta \mathrm{H})}{\Tr[\exp(- \beta \mathrm{H})]},
\end{equation}
 with inverse temperature $\beta=\log_{2}[p/(1-p)]$ and Hamiltonian $\mathrm{H}=\ket{1} \bra{1}$. That is because $\mathcal{T}_{q}$ preserves the diagonal states, while $\mathcal{R}_{p}$ transforms diagonal states into $\tau$. This defines a thermal state and a reference inverse temperature $\beta$. Note this is the same setting considered in Refs.~\cite{Guha2020,simonov2020ergotropy}, so we can properly compare our results.

In order to study quantum processes from the quantum thermodynamics perspective, we consider two figures of merit: free energy, as considered in Ref.~\cite{Guha2020}, and ergotropy, studied in Ref.~\cite{simonov2020ergotropy}.

The authors of Refs.~\cite{Guha2020,simonov2020ergotropy} showed that plugging the channels $\mathcal{R}_{p}$ and $\mathcal{T}_{q}$ into the quantum switch, thus implementing a quantum control of the order in which they act, results in a thermodynamic advantage with respect to all probabilistic mixtures of the channel compositions for the particular range of parameters considered in their study. That is, the quantum switch implies an increase in free energy or ergotropy when compared to the simple compositions of channels, or any probabilistic mixtures of the compositions, for the setting of parameters considered. Moreover, it has been claimed that -- for the specific range of parameters under analysis -- the indefinite causal order represented by the quantum switch operation results in a thermodynamic advantage~\cite{Guha2020,simonov2020ergotropy}. Here, we argue that indefinite causal order on its own does not result in thermodynamic advantages for quantum processes when considered a comparison of superchannels of the same form.

\subsection{Comparison of processes}\label{comparison}

\begin{figure}   
\begin{center}
\includegraphics[width=0.45\textwidth]{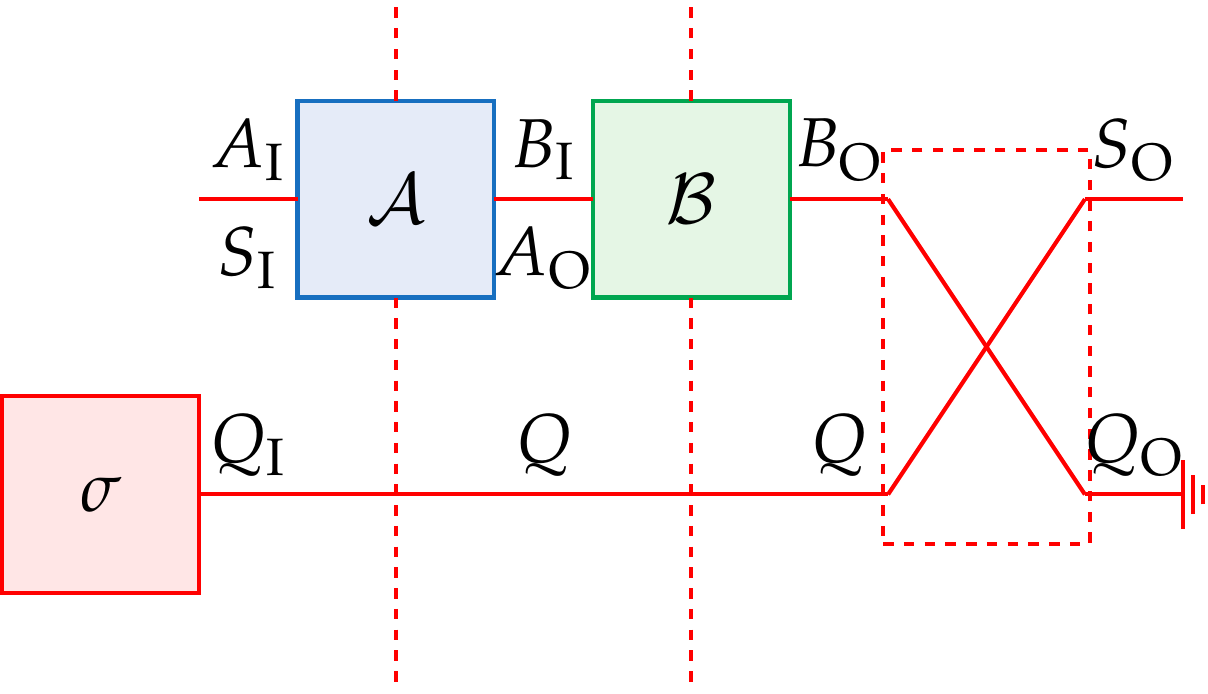}

    \caption{\textbf{Causally ordered process with arbitrary thermodynamic advantage}. Any two channels $\mathcal{A}$ and $\mathcal{B}$ (with appropriate dimensions of input and output spaces) are mapped to the replacement channel $\mathcal{C}(\mathrm{X})=\Tr[\mathrm{X}]\sigma$. This can be modelled as a {\sc swap} operation---represented in the dashed box---between the system and an environment prepared in state $\sigma$. As the state $\sigma$ can have arbitrary free energy (or any desired properties), this causally ordered process offers an arbitrary ``advantage'' as opposed to the simple sequential application of $\mathcal{A}$ and $\mathcal{B}$ on an input system.}
      \label{repl}
      
\end{center}
\end{figure}

The quantum switch is defined in Refs.~\cite{Guha2020,simonov2020ergotropy} with extra resources when compared to the composition of channels. Namely, the ancillary system responsible for the coherent control of orders in which the quantum operations are applied to the target system. Therefore, it is not entirely fair to compare the switch to the sequential application of operations, without the involvement of any additional system.

In fact, if we do not impose any restriction apart from definite causal order, one can always trivially increase the free energy (or the ergotropy) of a quantum system. As a simple example, let $\sigma$ be a quantum state with the desired property (free energy or ergotropy). We can define a causally ordered quantum process such that any input $(\mathcal{A},\mathcal{B})$ is mapped to the replacement channel $\mathcal{C}(\mathrm{X})=\Tr[\mathrm{X}]\sigma$. This can be realised physically by exchanging the target system with that of an uncorrelated ancilla prepared in state $\sigma$, see Fig.~\ref{repl}. Thus, independently of the action of the channels $\mathcal{A}$ and $\mathcal{B}$, and of the system's global past state, we can have arbitrary thermodynamic advantages defined by the fixed quantum state $\sigma$. 

\begin{figure}
\includegraphics[width=0.45\textwidth]{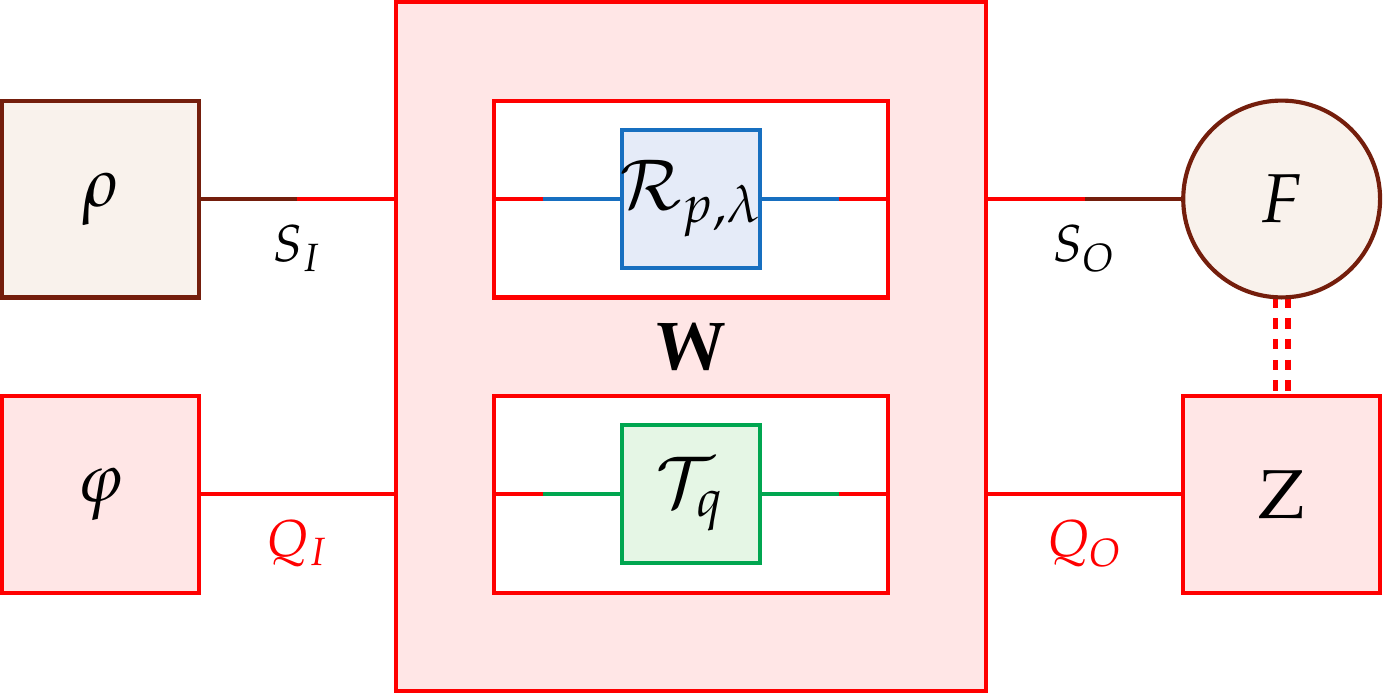}

    \caption{\textbf{Bipartite processes}. We consider only bipartite processes with bipartite global past and future. The global past state is uncorrelated, and the ancillary system is started in a pure state. A projective measurement is performed on the global future ancillary system. The free energy of the target system is averaged over the results of the measurement of the ancillary system.}
      \label{bipartiteprocess}
\end{figure}

This example might not be fully satisfactory, as the channels $\mathcal{A}, \mathcal{B}$ play no role in determining the final state, and the thermodynamic performance can be fully attributed to an independent system. This is quite different from the setup in Refs.~\cite{Guha2020,simonov2020ergotropy}, where the advantage is obtained after measuring an ancillary system and projecting the target on a particular state. In such protocols, ignoring the outcome of the measurement on the ancilla washes away the result. In order to provide a fair comparison between schemes with and without indefinite causal order, which have access to the same resources, we will restrict to the class of protocols depicted in Fig.~\ref{bipartiteprocess}, of which the setups in Refs.~\cite{Guha2020,simonov2020ergotropy} are particular cases.

In detail, such protocols start with a system $S_I$ and an ancilla $Q_I$ that are initially in a product state, $\rho \otimes \ket{\varphi}\bra{\varphi}$, where the ancilla's state $\ket{\varphi}$ is assumed to be pure. System and ancilla are then fed into the global past of a bipartite (or, as a generalisation, tripartite) quantum process.

After the action of the process on the channels, the global state of the output system $S_OQ_O$ is possibly entangled---depending upon the particular process. Then, a local projective measurement in a particular orthonormal basis $\{\ket{z_0},\ket{z_1}\}$ is performed on the ancillary system. The measurement is defined by the operators $\mathrm{M}_{j}=\mathbb{1}  \otimes \mathrm{Z}_{j}$, with $\mathrm{Z}_{j} = \ket{z_j} \bra{z_j}$ and $j=0,1$. Measuring the outcome $j \in \{0,1\}$ with probability (with similar definition for tripartite processes)
\begin{equation}
p_j=\Tr[\mathrm{M}_{j} \mathbf{W}[\mathcal{A},\mathcal{B}](\rho \otimes \ket{\varphi}\bra{\varphi})], \nonumber
\end{equation}
results in the conditional post-measurement state
\begin{equation}
	\sigma_{j} = \frac{1}{p_j} \mathrm{M}_{j} \mathbf{W}[\mathcal{A},\mathcal{B}](\rho \otimes \ket{\varphi}\bra{\varphi}) \mathrm{M}_{j}^{\dagger}.
	\label{ProjMeas}
\end{equation}

The relevant figures of merit are the average output free energy and ergotropy defined as
\begin{equation} \label{eq:thermoquantity}
T(\mathbf{W}) \coloneqq \sum_{j=0,1} p_j T(\sigma_i),
\end{equation} 
where $T$ generically denotes free energy $F$ or ergotropy $E$ of a quantum system. In general, the quantity defined in Eq.~\eqref{eq:thermoquantity} is a function on the initial state of $S_I$, input quantum channels $(\mathcal{A},\mathcal{B})$, and measurement operators $\{\mathrm{M}_0,\mathrm{M}_1\}$. 

\subsection{Non-Markovian processes as a thermodynamic resource} \label{results}

As we have seen, it is easy to reproduce an advantage in a thermodynamic process (increasing free energy or ergotropy) when an additional system, beyond the original one of interest, is available. The simplest example is the process shown in Fig.~\ref{repl}. No matter in which state we start, and regardless of the actions of the channels, the output target state is always $\sigma$, which can have arbitrary thermodynamic properties. 

We consider here a less trivial example, for which the external system interacts with the one of interest in a non-trivial way, and such that a final measurement on the external system and classical communication with the target is necessary to obtain an increase of free energy. This is closer in spirit to how the bipartite quantum switch was used to extract free energy and ergotropy in Refs.~\cite{Guha2020,simonov2020ergotropy}.

Let our quantum system of interest $S_I$ interact with an ancillary qubit system $Q_I$ through the Ising Hamiltonian 
\begin{equation}
\mathrm{H}_{\mathrm{Ising}}=-\sigma_{x} \otimes \sigma_{x} -(\mathbb{1}  \otimes \sigma_{z}+\sigma_{z}\otimes\mathbb{1} ), \nonumber
\end{equation}
where $\sigma_{x}=\ket{0}\bra{1}+\ket{1}\bra{0}$ is a Pauli matrix.

That defines a bipartite unitary quantum channel
\begin{equation} 
\mathcal{U}(\rho)=\mathrm{U}_{\mathrm{Ising}} \rho \mathrm{U}_{\mathrm{Ising}}^{\dagger}, \nonumber
\end{equation}
with $\mathrm{U}_{\mathrm{Ising}} \coloneqq \exp \left( -\ii \mathrm{H}_{\mathrm{Ising}} \right)$. We set the Plank constant $\hbar$ and time displacement equal to unit in the evolution operation $\mathrm{U}_{\mathrm{Ising}}$. 

\begin{figure}   
\includegraphics[width=0.45\textwidth]{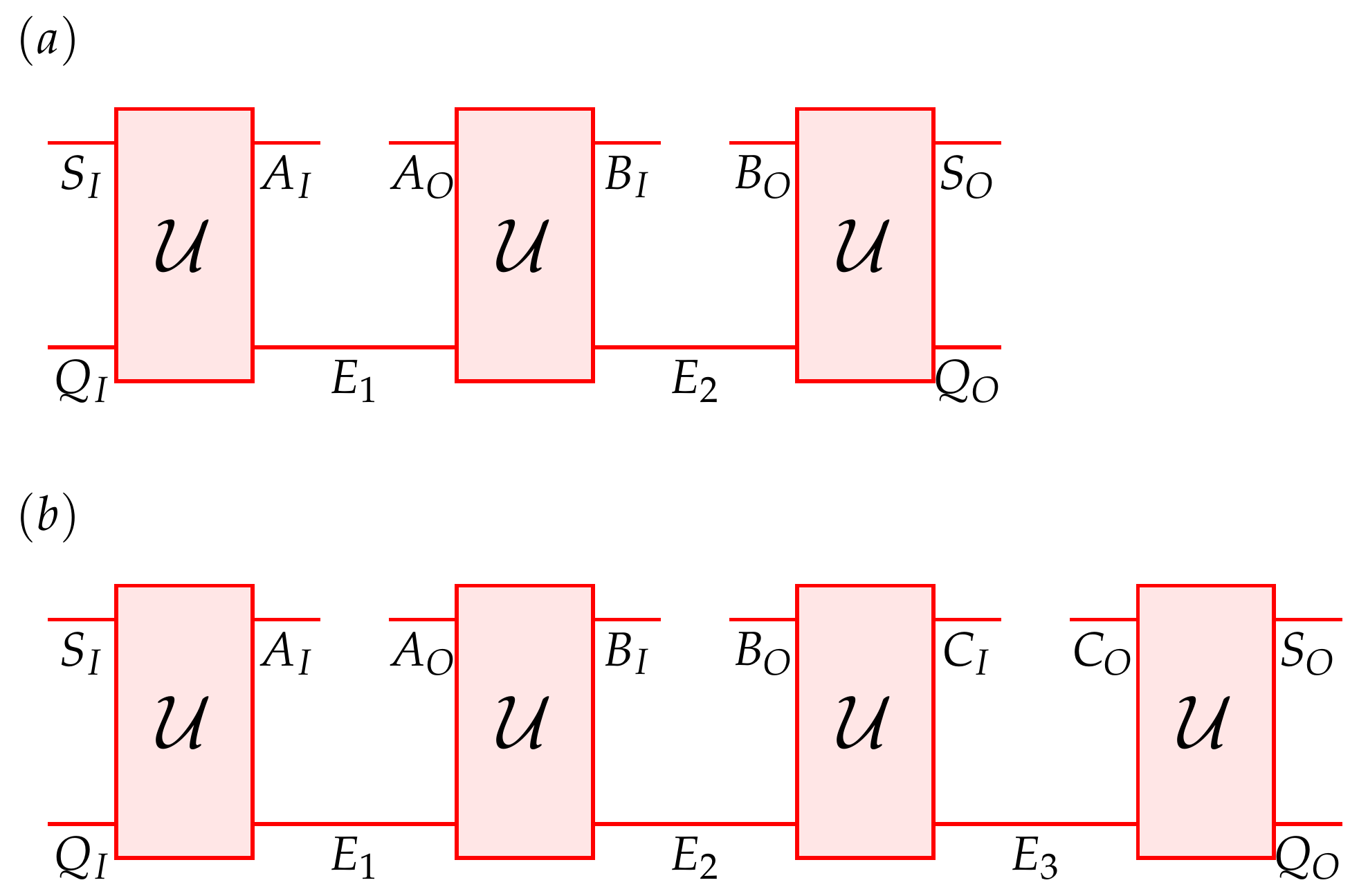}

    \caption{\textbf{Non-Markov processes defined through Ising model interactions.} The bipartite and tripartite processes defined in Eqs.~\eqref{eq:bipartite_Ising} and \eqref{eq:tripartite_Ising} are represented in top and botton panels, respectively. Here, we consider the ancillary system $Q_I$ in the pure quantum state $\ket{+}$, and perform a projective measurement on $Q_O$.}
      \label{fig:ISINGNMdiagram}
\end{figure}

Let us define the bipartite non-Markovian process
\begin{equation} \label{eq:bipartite_Ising}
\mathbf{W}_{\mathrm{Ising}}^{(2)} \left( \mathcal{R},\mathcal{T} \right) \coloneqq  \mathcal{U} \circ ( \mathcal{T}\otimes \mathcal{I}) \circ \mathcal{U}\circ ( \mathcal{R}\otimes \mathcal{I}) \circ \mathcal{U}, 
\end{equation}
where the identity map acts on the ancillary system. The top panel of Fig.~\ref{fig:ISINGNMdiagram} shows a diagram for this process.

Analogously, we define the tripartite non-Markovian process (see the bottom panel in Fig.~\ref{fig:ISINGNMdiagram})
\begin{multline} \label{eq:tripartite_Ising}
\mathbf{W}_{\mathrm{Ising}}^{(3)} \left( \mathcal{R},\mathcal{R},\mathcal{T} \right) \coloneqq \\ \mathcal{U} \circ ( \mathcal{T}\otimes \mathcal{I}) \circ \mathcal{U} \circ ( \mathcal{R}\otimes \mathcal{I}) \circ \mathcal{U} \circ ( \mathcal{R}\otimes \mathcal{I}) \circ \mathcal{U}.
\end{multline}

For the case of bipartite processes we compare the following examples.
 
\begin{itemize}
    \item \textbf{Composition of channels}. We consider the consecutive application of the channels $\mathcal{R}_{p}$ and $ \mathcal{T}_{q}$. 
    \item \textbf{Bipartite quantum switch $\mathbf{W}_{\textrm{2-switch}}$}. We consider $(\mathcal{R}_{p},\mathcal{T}_{q})$ as input channels. The control system is initially prepared in the state $\ket{+}$, thus, providing maximal control of the causal orders.
    \item \textbf{Bipartite non-Markovian process $\mathbf{W}_{\mathrm{Ising}}^{(2)}$}. We consider the input channels $(\mathcal{R}_{p},\mathcal{T}_{q})$. The ancillary qubit is prepared in the state $\ket{+}$.
\end{itemize}  

Considering tripartite processes, we compare the following examples.

\begin{itemize}
    \item \textbf{Composition of channels}. We consider the consecutive application of two GAD and one PF channel. 
    \item \textbf{Tripartite quantum switch $\mathbf{W}_{\textrm{3-switch}}$}. We consider the input channels $(\mathcal{R}_{p},\mathcal{R}_{p},\mathcal{T}_{q})$.  The orders considered are $A\prec B\prec C$ and $C\prec B\prec A$. The control system is initially prepared in the state $\ket{+}$.
    \item \textbf{Tripartite pure quantum process $\mathbf{W}_{\textrm{Lugano}}$.}  We consider the input channels $(\mathcal{R}_{p},\mathcal{R}_{p},\mathcal{T}_{q})$.  The second ancillary system is defined with input in the quantum state $\ket{0}$ ($\ket{+}$) for free energy (ergotropy) calculations and then the output system is discarded, thus, implementing a tripartite quantum superchannel with bipartite global past and future. 
    \item \textbf{Tripartite non-Markovian process $\mathbf{W}_{\mathrm{Ising}}^{(3)}$.} We consider the input channels $(\mathcal{R}_{p},\mathcal{R}_{p},\mathcal{T}_{q})$. The ancillary system is started in the quantum state $\ket{+}$. 
    \label{list1}
\end{itemize}

\subsubsection{Free energy}

We consider here the free energy of a qubit system $S$ after the action of different quantum processes. We take a system with initial and final Hamiltonian $\mathrm{H}=\ket{1}\bra{1}$. With respect to a thermal reservoir at inverse temperature $\beta$, we define the free energy of the quantum system $S$ in the state $\rho$ as \cite{horodecki2013fundamental,brandao2013resource}
\begin{equation}
F_{\beta}(\rho)=\Tr[\mathrm{H}\rho]-\beta^{-1}\ln(2)S(\rho), \nonumber
\end{equation}
where $S(\rho) \coloneqq - \Tr[\rho \log_{2}\rho]$ is the von Neumann entropy of a quantum state.

\begin{figure}   
\begin{adjustbox}{ width= \linewidth}
\centering 
\includegraphics[]{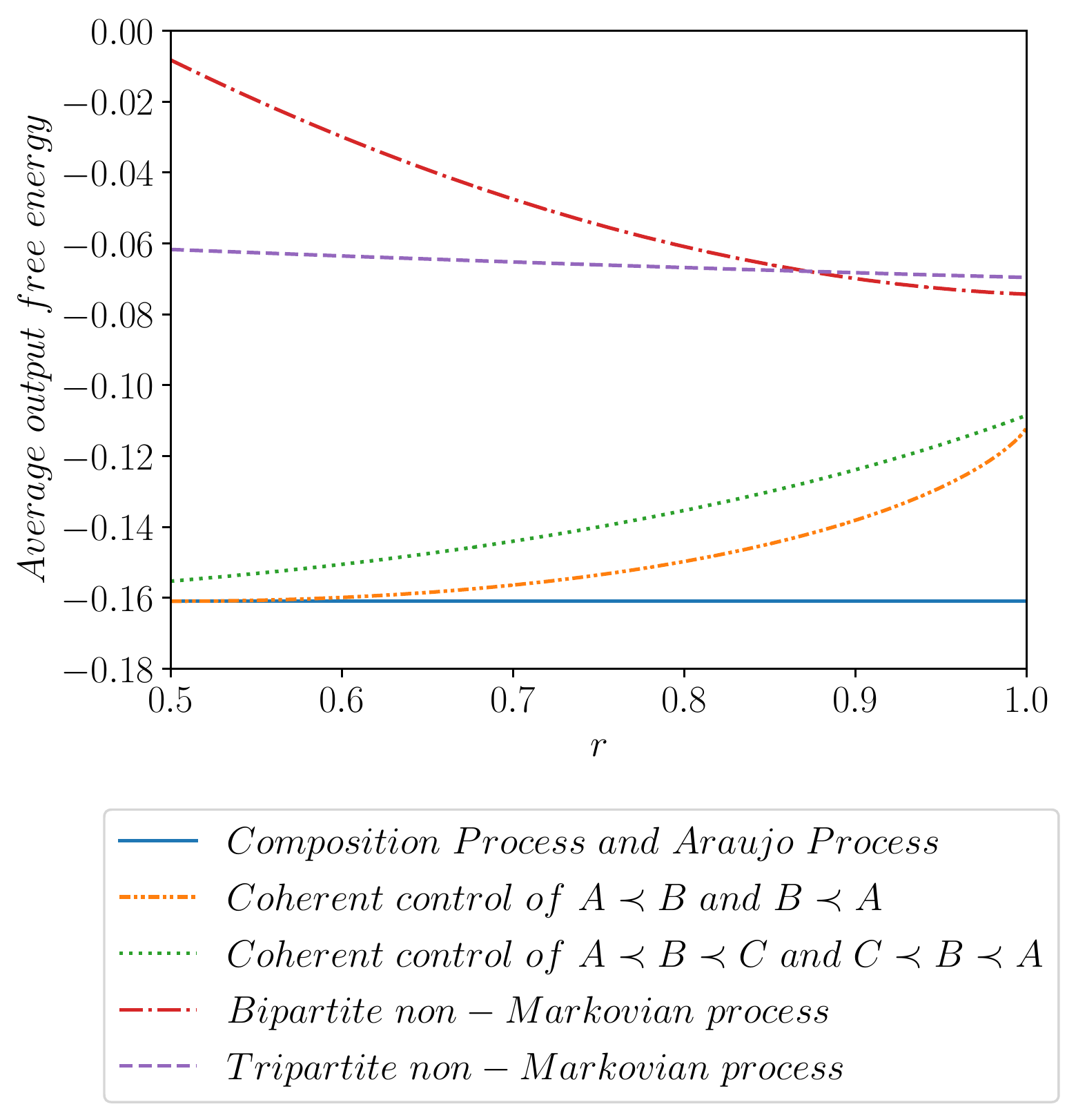} 
\end{adjustbox}
    \caption{\textbf{Free energy for diferent quantum processes}. The global past state is the diagonal state $\rho=r\ket{0}\bra{0}+(1-r)\ket{1}\bra{1}$. The ancillary global past state is $\varphi=\ket{+}\bra{+}$, with $\ket{\pm}=(\ket{0} \pm \ket{1})/\sqrt{2}$. The local projective measurement on the ancillary global future state is defined by the basis $\{\ket{+},\ket{-}\}$. The channel parameters considered here are $\lambda=1$, $p=q=0.8$.}
      \label{fig:plot1}
\end{figure}

Let us start considering bipartite quantum processes. Figure~\ref{fig:plot1} shows that we can use the non-trivial interaction between $S_O$ and $Q_O$ in order to increase the output free energy. As reported in Ref.~\cite{Guha2020}, the quantum switch results in an associated average free energy which is higher than the one obtained in the case of the channel composition, for any value $r >0$. This allows work extraction from the output system $S_O$, which is not possible with the channel composition solely. Nevertheless, this result is not exclusive for processes with indefinite causal order. The non-Markovian process defined in Eq.~\eqref{eq:bipartite_Ising} has similar behaviour. In fact, it results in higher average free energy when compared with the bipartite quantum switch for any value of $r$. Furthermore, this examples show that indefinite causal order is not a necessary condition on the thermodynamic advantages considered here. 

Now, let us consider the case of tripartite processes with bipartite global past and future. Similar conclusions with respect to the case of bipartite processes can be made. That is, the tripartite causally ordered non-Markovian process in Eq.~\eqref{eq:tripartite_Ising} implies a greater average output free energy when compared with the tripartite quantum switch defined in Eq.~\eqref{eq:tripartite_switch}.

Now, we examine the existence of a tripartite process with indefinite causal order, but with no thermodynamic advantage. We can use the process defined in Eq.~\eqref{eq:det_process} in order to build a tripartite process with bipartite global past and future. That can be achieved by feeding the overall channel $\mathbf{W}_{\mathrm{det}}(\mathcal{A},\mathcal{B},\mathcal{C})$ with an input state $\ket{0}\bra{0}$ of the system ${C_{I}^{(3)}}$ and tracing out the resulting output system $C_{O}^{(3)}$. In the following discussion we also denote this process with the symbol $\mathbf{W}_{\mathrm{det}}$. Figure~\ref{fig:plot1} shows that for the parameters considered, $p=q=0.8$, the process $\mathbf{W}_{\mathrm{det}}$ do not imply in a thermodynamic advantage when compared with the composition of channels. Thus, indefinite causal order solely is not a sufficient condition on the advantage addressed here.

\subsubsection{Ergotropy}

In contrast to free energy, representing the maximum work that can be extracted from the system while it is in contact with a thermal bath (i.e. at a constant temperature), \textit{ergotropy} refers to work extraction from a system under cyclic unitaries. In principle, certain unitaries can be used for maximum work extraction which transform the initial state of the system into a thermal state, called \textit{completely passive}, from which no work can be extracted. Therefore, this quantity ---\textit{ergotropy}--- provides a bound on the extractable work from a system using cyclic unitaries. The formal definition, for an initial state $\rho$ is~\cite{allahverdyan2004maximal}
\begin{equation}
    E(\rho)=\max _{\mathrm{U}} \Tr \left[\mathrm{H} (\rho-\mathrm{U} \rho \mathrm{U}^{\dagger})\right]~.
    \label{eq:ergo}
\end{equation}

Here, the cyclic unitary process can be achieved by a time-dependent Hamiltonian $\mathrm{H}$ which drives the system under the restriction $\mathrm{H}(t)= \mathrm{H}(0)$, where $t$ is the driving period. During the driving period, $\mathrm{H}(s) = \mathrm{H}(0) + \mathrm{V}(s)$, where $\mathrm{V}(s)$ can be identified as the driving Hamiltonian for $0 < s < t$. While calculating the ergotropy of a system, by definition, requires a maximization over all possible unitaries, here we have considered a single qubit system for which the ergotropy can be analytically computed from its state $\rho$. 

In general, the formalism of work extraction using cyclic unitary processes can also be extended to quantum channels where the system is governed by an open quantum dynamics during the driving period. Here, we consider the initial and final states of a qubit, which undergoes transformation through the quantum channels based on quantum processes, as discussed in the previous section. The ergotropy vested in the initial and final states of the qubit can be compared and the changes, if any, can be calculated.

Another important concept that we have used extensively henceforth is that of \textit{daemonic ergotropy}~\cite{francica2017daemonic}. This is the ergotropy obtained upon the projective measurement of the ancilla ($A$) such that the post measurement outcome gives some information about the target system ($S$). The gain due to access to this information, by virtue of correlation between ancilla and the target is called \textit{daemonic gain} on account of its similarity to the Maxwell's demon, where thermodynamic work is enhanced by access to extra information about the state of the system. In order to defining the \textit{daemonic ergotropy}, the probabilities associated with the measurement on the ancilla are required, which are given by
\begin{align}
    p_\alpha = \operatorname{Tr}\{\pi_\alpha \rho_{SA}\}~,
\end{align}
where $\alpha$ labels the measurement outcome obtained through projective measurement operator $\pi_\alpha$ on the joint state of system and ancilla, given as $\rho_{SA}$.
Upon the post selection of the measurement outcome of the ancilla, the state of the target qubit (of which the ergotropy is measured) is given by
\begin{align}
    \rho_{S|A}^\alpha = \frac{1}{p_\alpha}\operatorname{Tr}_A\{\pi_\alpha \rho_{SA}\}
\end{align}
The average daemonic ergotropy can be computed as
\begin{align}
E^D_{\pi_\alpha} = \sum_\alpha p_\alpha E(\rho_{S|A}^\alpha)
\label{eq:avgdergo}
\end{align}
It is important to note that the states $\rho_{SA}$ and $\rho_{S|A}$ that we consider here are the joint output states and the conditional state of the target, respectively, which are obtained after the application of channels based on quantum processes. The post selection of the ancilla, therefore, ensures that an appropriate evolution is selected for the target qubit, which results in a certain ergotropy. Since the task is to maximise the energetic content of the (post selected) state 
(see Eq.~\eqref{eq:ergo}), it translates to finding the appropriate orthogonal basis $\{\pi_\alpha\}$ of the measurement on the ancilla, which maximises the daemonic ergotropy in Eq.~(\ref{eq:avgdergo}).

Here, the generalized measurement bases (orthogonal) considered for the ancillae in different cases are
\begin{align}
    |M_1\rangle &= \sqrt{p}|0 \rangle + e^{i\phi}\sqrt{1-p}|1\rangle ~, \nonumber\\ 
    |M_2\rangle &= -e^{i\phi}\sqrt{1-p}|0\rangle  + \sqrt{p} |1\rangle~.
    \label{eq:mbasis}
\end{align}
This measurement basis is used for the control qubit in the case of the switch (bipartite/tripartite) and the ancilla in the process $\mathbf{W}_{\mathrm{det}}$, and the (bipartite/tripartite) non-Markovian processes. The probabilities obtained after the measurement in this basis are used as weights for the ergotropy of the target (which is calculated analytically from the reduced state of the target system, after the measurement of the control), as outlined above in Eq.~\eqref{eq:avgdergo}.

Additionally, we also allow an added flexibility on the initial (input) state of the ancilla/control. This is achieved with $\rho_A = \cos{(x/2)} |0\rangle\langle 0 | + e^{i\chi} \sin{(x/2)} \ket{1}\bra{1}$, where $x\in [0,\pi]$, and $\chi \in [0,2\pi]$. Therefore, in all the cases that we consider, the optimization is also performed over the input state of the ancilla, i.e., the parameters $x$ and $\chi$, in addition to the measurement basis in Eq.~\eqref{eq:mbasis}.

Effectively, the task reduces to the optimization over the parameters $p$, $x$, $\chi$ and $\phi$ for maximum daemonic ergotropy in Eq.~\eqref{eq:avgdergo}. We perform this optimization using the basinhopping routine in Scipy, with sequential least square programming (SLSQP) as the core algorithm. As a benchmark, we consider the bipartite quantum switch setup which has been considered previously in Ref.~\cite{simonov2020ergotropy}, and find that this algorithm is capable of recovering the optimized parameters found analytically therein. 

In the previous literature~\cite{simonov2020ergotropy}, the bipartite quantum switch configuration with GAD and PF has been shown to provide an advantage in ergotropy over the composition of channels, with a particular choice of measurement basis of the ancilla (control). There, it was established that the maximum Daemonic ergotropy is found when the control is prepared in the $|+\rangle$ basis and subsequently measured in the orthogonal basis $\{|+\rangle, |-\rangle\}$ on the output side. Consequently, the daemonic ergotropy is $E^D_{\pi_{\ket{\pm}}} = p_+ E(\rho_{S\ket{+}}) + p_- E(\rho_{S\ket{-}})$, wherein the ergotropies $E(\rho_{S\ket{\pm}})$ can be calculated analytically from the post selected states $\rho_{S\ket{\pm}}$. In the aforementioned work, the input target state were $\sqrt{r}|0\rangle+\sqrt{1-r}|1\rangle$ while and the channel parameters were $p = 1/3$, $q = 0$ and $\lambda = 1/2$. The gain in ergotropy, compared over population imbalance of the input state ---$\delta \rho \coloneqq \rho_{22}-\rho_{11}$, where $\rho_{22},~\rho_{11}$ are the diagonal entries of $\rho$--- had been attributed to the coherent control over the order of applications of the maps. Here we expand the paradigm of process-induced ergotropic gains to include the Lugano process and also a non-markovian processes. We choose the same channel parameters as well as the parametric initial state of the target, and also base our results on the population imbalance in the initial state $\delta \rho$.

\begin{figure}
\includegraphics*[width=1\linewidth]{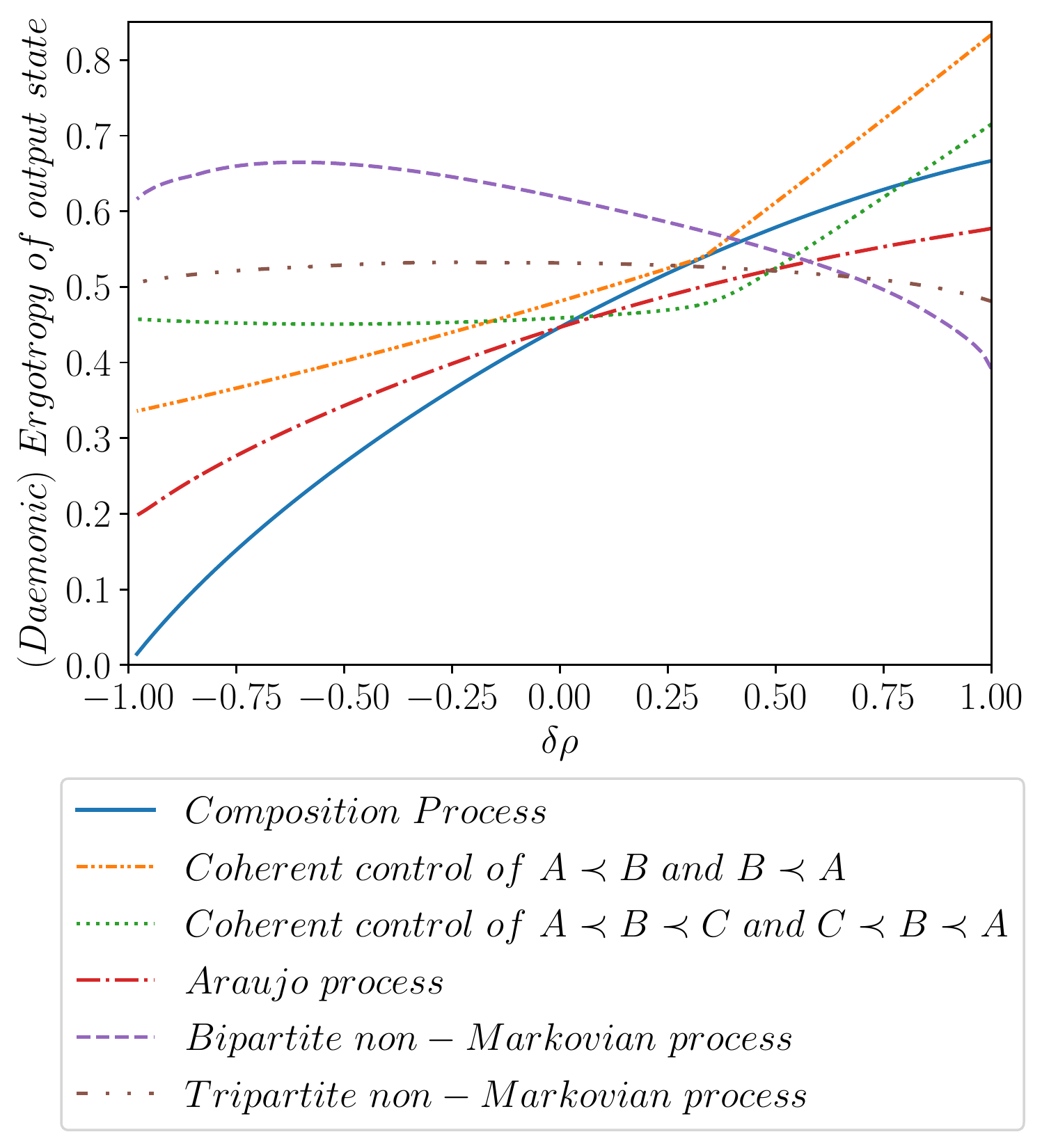}
 \caption{Maximum ergotropy obtained after optimizing over preparation and measurement of ancilla, as a function of population imbalance $\delta \rho$. The input state of the target considered in all the cases is $\big[\sqrt{r}|0\rangle+\sqrt{1-r}|1\rangle \big]$. The channel parameters considered are  $p = 1/3,~q = 0,~\lambda = 1/2 $. Since there are two ancillae in the Luganoprocess, one of them is discarded at the output while the other is optimized over in measurement basis. This is done for a fair comparison between all of the cases, where only one ancilla is considered uniformly in all the other setups.}
 \label{plot2}
\end{figure}

In Fig.~\ref{plot2}, we show that for the non-Markovian processes, the maximized daemonic ergotropy over ancilla preparation and measurement could be more than the bi- and tri-partite switch, $\mathrm{det}$ process, and the separable configuration of channels' composition. Though the optimized parameters defining the ancilla preparation/measurement bases in all the cases are interesting in their own right, we have skipped their discussion to focus only on the ergotropic gains. We also show in Fig.~\ref{plot2} that for the tri-partite process $\mathbf{W}_{\textrm{Lugano}}$ (with the same channel parameters and input target state), the ergotropy may or may not be more than that in the composition process. We note that the process $\mathbf{W}_{\textrm{Lugano}}$ can violate causal inequalities \cite{baumeler2016space}, so it has an even stronger form of indefinite causal order than the quantum switch (which cannot). However, even such a powerful resource does not relate directly to advantages on the thermodynamic settings considered here. Through the trends obtained in various cases, these results void the possibility of establishing a simple connection between the nature of quantum processes and the daemonic gain in ergotropy, possible through preparation and measurement of ancilla in an optimal scenario. 

\section{Conclusions} \label{conclusions}

We considered the role of indefinite causal order in the context of quantum thermodynamics. The results presented in Refs.~\cite{Guha2020,simonov2020ergotropy} claimed that indefinite causal order of the switch provides an advantage in the output free energy and ergotropy, respectively, when compared with the composition processes for a fixed strategy with the appropriate range of parameters. We claim here that, under the reasonable comparison of processes of the same form, and with the assistance of the same type of resources, indefinite causal order offers no clear advantage for the above mentioned tasks.

Specifically, we studied the free energy and ergotropy generated by the action of higher order transformations on channels of interest (GAD and PF) and compared processes with definite and indefinite causal order having access to comparable resources---namely, an additional ancillary system. Allowing for measurements on the ancillary systems, we found that the causally ordered processes generally perform similarly or better than those with indefinite causal order. A reasonable interpretation is that the advantages are due to the non-trivial interaction between system and ancilla, rather than to indefinite causal order, as the latter does not necessarily imply higher output free energy and ergotropy. We have considered here examples of bipartite and tripartite processes supporting this claim. Although indefinite causal order is shown not to be an advantage for quantum thermodynamics in the tasks considered here, we do not discard this possibility for other protocols. Finding protocols for which an advantage is associated with indefinite causal order is left for future studies. Indeed, studies conducted in this direction have been successful for quantum information problems~\cite{quintino2019reversing,quintino2022deterministic}.

We can further develop such a reasoning. In order to talk about thermodynamic advantages arIsing from indefinite causal order, we must first define a measure of such a property. One could, for example, attempt to link thermodynamic advantages with some measure of indefinite causal order (such as the generalized robustness introduced in Ref. \cite{Araujo2015}). Let us consider a situation where, under certain set $X$ of conditions, the free energy (or some other thermodynamic property) is less than a given value $T_0$ as long as we have definite order. In this case, if we have the considered property larger than $T_0$, and the set $X$ is satisfied, then we can conclude that there was indefinite causal order. In other words, $T(\mathbf{W})>T_0 \Rightarrow f(\mathbf{W})>0$, for some real-valued function $f$ on the space of bipartite processes with bipartite global past and future, quantifying the extent to which the process is not causally separable.

Our examples show that this is not the case for the tasks considered here. Any reasonable causal non-separability measure $f$ would necessarily satisfy $f(\mathbf{W}_{\textrm{switch}}) > f(\mathbf{W})$, for any causally ordered process $\mathbf{W}$. Strict inequality must hold since the quantum switch in not causally ordered. Thus the condition given above is not satisfied in the examples considered in our work, as can be seen in Figs.~\ref{fig:plot1} and~\ref{plot2}.

A distinct perspective rests on the evaluation of the thermodynamic cost for implementing quantum processes, which remains an open problem. In fact, results in this direction have been presented in recent literature for situations not precisely equivalent to the one considered in this paper. For instance, the authors in Ref.~\cite{faist2019thermodynamic,faist2021thermodynamic} defined the thermodynamic capacity of quantum channels as the work cost for implementing quantum channels in the asymptotic regime. The former stands as an extension for quantum channels of previous results valid for quantum states~\cite{brandao2013resource}. The approach taken into account there is that of resource theories, and furthermore, stands as a promising avenue for defining the work cost of implementing higher-order operations as well. A different and promising approach has been conducted in the Ref.~\cite{liu2022thermodynamics}, where the implementation of the quantum switch was associated to energy costs. Nevertheless, this consideration is left as an open problem. Therefore, novel studies must be performed in order to address such a problem.

\begin{acknowledgments}
MC and LCC acknowledge the financial support from funding agencies CNPq, FAPEG and the Brazilian National Institute of Science and Technology of Quantum Information (INCT- IQ). This study was financed in part by the Coordenação de Aperfeiçoamento de Pessoal de Nível Superior – Brasil (CAPES) – Finance Code 001. MC also acknowledges the warm hospitality of the School of Mathematics and Physics at the University of Queensland, where this study was developed. FC acknowledges support through an Australian Research Council (ARC) Discovery Early Career Researcher Award (DE170100712) and by the ARC Centre of Excellence for Engineered Quantum Systems (Project
No.\ CE17010000). The University of Queensland (UQ) acknowledges the Traditional Owners and their custodianship of the lands on which UQ operates. The authors acknowledge the financial support received from the IKUR Strategy under the collaboration agreement between Ikerbasque Foundation and BCAM on behalf of the Department of Education of the Basque Government.
\end{acknowledgments}

\end{document}